\documentclass[10pt]{article}
\usepackage[dvips]{epsfig}
\usepackage[T1]{fontenc}
\usepackage[latin1]{inputenc}
\usepackage{graphicx}
\usepackage[english]{babel}
\usepackage{amsmath}
\usepackage{amssymb}
\usepackage{amsfonts}
\usepackage[T1]{fontenc}
\usepackage{color}
\usepackage{babel}
\usepackage{verbatim}
\usepackage[unicode=true,pdfusetitle,bookmarks=true,bookmarksnumbered=false,bookmarksopen=false,
 breaklinks=false,pdfborder={0 0 1},backref=false,colorlinks=true]{hyperref}
\hypersetup{linkcolor=blue,citecolor=blue}
\makeatletter
\usepackage[dvips]{epsfig}
\usepackage[T1]{fontenc}
\usepackage{color}
\usepackage{pdflscape}

\begin{document}

\title{\bf On the initial singularity in  Kantowski-Sachs spacetime}

\author{Elham Ghorani$^{1}$\thanks{%
email: e.qorani@gmail.com},~and~Yaghoub Heydarzade$^{2}$\thanks{%
email: yheydarzade@bilkent.edu.tr}
\\{\small $^1$Gorgan, Iran}\\
\small{$^2$Department of Mathematics, Faculty of Sciences, Bilkent University, 06800 Ankara, Turkey}}
\date{\today}

\maketitle
\begin{abstract}
The emergent universe scenario is a proposal for resolving the Big Bang singularity
problem in the standard Friedmann-Lemaitre-Robertson-Walker  cosmology. In the context of this scenario, the Universe originates from a nonsingular static state.  In the present work, considering the realization of the emergent universe
scenario, we address the possibility of having a nonsingular Kantowski-Sachs type static state.  Considering four and five dimensional models (with and
without brane), it is  shown that both the existence and stability of a nonsingular state depend on the  dimensions of the spacetime and the nature of the fluid supporting the geometry.
\end{abstract}
\maketitle

%%%%%%%%%%%%%%%%%%%%%%%%%%%%%%%%%%%%%%%%%%%%%%%%%%%%%%%%%%%%%%%%%%%%%%%%%%%%%%%%%%%%%%%%%%%%%%%%%

\section{Introduction}
Besides the great successes of standard model of cosmology based on Einstein's
theory of general relativity (GR), there  still remain several theoretical issues.
One attempt to resolve these problems was the inflationary universe scenario \cite{inflation}.
Although the inflation resolves some of the problems such as horizon problem,
flatness problem and magnetic monopole problem, but it is incapable of solving
the initial big bang singularity  problem. In  the standard model of cosmology, the Universe originates from an initial
singular point where all the mass, energy, and spacetime are infinitely compressed. To cure this singularity problem, some models such as the ekpyrotic/cyclic
universe \cite{cyclic}, the pre-big bang  \cite{prebigbang}, and the emergent universe \cite{emergent} have been
proposed. In the latter scenario proposed by Ellis et al \cite{emergent},  the Universe
has no timelike singularity, it is ever existing and stands
almost static in the infinite
past. The emergent model replaces the initial big bang singularity with  a nonsingular
static state,  the so called Einstein static state (ESS). Then, the Universe enters to an inflationary
era and
produces the same cosmological history as in the standard model.
The existence
of a stable ESS against various perturbations, such as quantum fluctuations, is a prerequisite  for a successful
emergent universe scenario. The original emergent model supposes an ESS that
is characterized by
Friedmann-Lemaitre-Robertson-Walker (FLRW) metric
with a perfect fluid source. In the framework of  Einstein's GR, it was demonstrated
by Eddington   that an ESS  is unstable versus the homogeneous and isotropic
perturbations \cite{perturb}.
Later studies by
Gibbons \cite{gib} and Barrow et al \cite{barr} showed that  an ESS  can be neutrally stable
versus small inhomogeneous vector
and tensor perturbations as well as adiabatic scalar density
perturbations  if the perfect fluid filling the universe has a sound speed $c_s^2 > 1/5$.
 An approach to amend the instability problem is to modify cosmological field
equations of GR. Many works has been performed
along this line in the context of modified theories of gravity   such as  Einstein-Cartan theory \cite{ec}, $f(R)$ gravity \cite{fr}, $f(T)$ gravity \cite{ft},  brane
gravity \cite{brane}, massive gravity \cite{mass} and modified Gauss-Bonnet gravity \cite{gb} among the others.

The emergent universe scenario has been analyzed in Einstein's GR and modified
theories only for the FLRW spacetime. One interesting question that can be raised
here is: \textit{Can a non-FLRW static state be a seed for an emergent universe?} To answer this question, one may consider  Kantowski-Sachs (KS) \cite{kan} or Bianchi  type cosmological models \cite{bian} and investigate their stability
versus various perturbations. Our aim in the present work is to explore the former.
 KS
models  are spatially homogeneous and rotation-free but possess shear. Hence,
they represent anisotropic universes. However, it is shown
that   the initial anisotropies in the context of these models die away as the
universe expands from the initial singularity if there is a positive cosmological constant, which is
effectively the requirement for an inflation scenario in the early stages of our universe \cite{die1, die2}. Therefore, the universe could be initially anisotropic  KS type, and then it passes through the inflation and its subsequent cosmological eras  toward the current homogeneous and  isotropic state.
In \cite{die1},  KS type anisotropic cosmological models  are classified in the presence of a nonzero cosmological constant. It is
shown that for a positive cosmological constant  there exists a set of big
bang models of zero measure as well as a set of models possessing nonzero measure with the de Sitter asymptotic.  In \cite{noj}, nonsingular KS type
cosmological solutions are obtained in a quantum-corrected Einstein gravity.
In deriving these analytical solutions, the authors used the analogy with the Nariai black hole. The global structure of the KS cosmological models
are studied in \cite{global}. It has been shown that if the energy-momentum source is a perfect fluid, all the general relativistic models are geodesically incomplete, both to the past and to the future, and
the energy density of the perfect fluid diverges   at each resulting singularity.
In \cite{cosm},  KS models are examined versus the classical tests of cosmology,  and the results have been compared with those belonging to FLRW spacetime in the standard model of cosmology. It is shown that for a large class of KS models, the observations are incapable of distinguishing between KS models and the standard model of cosmology.  The reference \cite{pert} discusses the
growth of density perturbations in KS cosmologies  in the presence of a positive cosmological constant. It is found that when a  bounce occurs in the cosmic scale factor, the density gradient in the bouncing directions experiences a local maximum at or slightly after the bounce.

The organization of this paper is as follows. In section 2, we introduce
the Kantowski-Sachs metric and its corresponding Einstein field equations
in four dimensions.
We obtain the existence and stability conditions versus the scalar perturbations
for a Kantowski-Sachs static state
(KSSS)  for various
energy-momentum
sources with  generic  linear and Chaplygin gas type equations of
state. The analysis in four dimensions shows that there is only one particular fluid type that can support a nonsingular KSSS.
In section 3, regarding the result in section 2 and motivated by the fact that anisotropic models can be exact solutions to the string effective action \cite{string}, we apply our analysis to a braneworld generalization of the
model
 in which a four dimensional
KS spacetime  embedded in a five dimensional bulk space. The  conditions for the existence and stability of a nonsingular  KSSS are also discussed in this framework for various
 fluid types. In section 4, the stability analysis is done for a 5-dimensional KS model without the brane.  Finally, section 5 is devoted to our concluding remarks.
%%%%%%%%%%%%%%%%%%%%%%%%%%%%%%%%%%%%%%%%%%%%%%%%%%%%%%%%%%%%%%%%%%
%%%%%%%%%%%%%%%%%%%%%%%%%%%%%%%%%%%%%%%%%%%%%%%%%%%%%%%%%%%%%%%%%%%%%%%%%
%%%%%%%%%%%%%%%%%%%%%%%%%%%%%%%%%%%%%%%%%%%%%%%%%%%%%%%%%%%%%%%%%%%%%%%%
\section{KS geometry in four dimensions  }
We  consider that the spacetime metric  is the KS type \cite{kan}
\begin{equation}\label{11}
ds^2=-dt^{2}+a_{1}^{2}(t)dr^{2}+a_{2}^{2}(t)(d\theta^{2}+sin^{2}\theta
d\varphi^{2}),
\end{equation}
where $a_{1}(t)$ and $a_{2}(t)$ are two arbitrary functions of time
and the energy-momentum tensor supporting this geometry has the
generic form ${T^{\mu}}_{\nu}=diag\left(-\rho,\,p_r,\,p_t,
\,p_t\right)$ \cite{Ram}. Then, the Einstein field equations
considering the above metric and energy-momentum source read as
\begin{eqnarray}
&&\frac{\dot a_{2}^{2}}{a_{2}^{2}}+\frac{2\dot a_{1}\dot
a_{2}}{a_{1}a_{2}}+\frac{1}{a_{2}^{2}}=\Lambda+ k_{4}\rho,\label{fe1}\\
&&\frac{2\ddot a_{2}}{a_{2}}+\frac{\dot
a_{2}^{2}}{a_{2}^{2}}+\frac{1}{a_{2}^{2}}=\Lambda- k_{4} p_{r},\label{fe2}\\
&&\frac{\ddot a_{1}}{a_{1}}+\frac{\ddot a_{2}}{a_{2}}+\frac{\dot
a_{1}\dot a_{2}}{a_{1}a_{2}}=\Lambda- k_{4}p_{t}\label{fe3},
\end{eqnarray}
where $k_4=8\pi G$ is the four dimensional gravitational coupling
constant  and $\Lambda$ is a generic (positive or negative) cosmological constant.
\subsection{KSSS and stability analysis
in four dimensions} In the following, we investigate the existence
and stability of KS type static state considering the field
equations (\ref{fe1})-(\ref{fe3}). To keep the generality of the
analysis, we consider two generic kinds of energy-momentum sources: $(i)$ a fluid possessing linear equation of state, and $(ii)$ a fluid with generalized Chaplygin gas
type equations of state.
\subsubsection{Energy-momentum source with
linear equations of state} For the sake of generality, here we consider
distinct equations of state for the radial and lateral pressures
$p_r$ and $p_t$, respectively, as \cite{Ram, Eros, Khad}
\begin{equation}\label{eos1}
p_{r}=\omega_{r}\rho +p_{0r},~~~~p_{t}=\omega_{t}\rho +p_{0t}.
\end{equation}
Then, the field equations (\ref{fe1})-(\ref{fe3}) become
\begin{eqnarray}
&&\frac{\dot a_{2}^{2}}{a_{2}^{2}}+\frac{2\dot a_{1}\dot
a_{2}}{a_{1}a_{2}}+\frac{1}{a_{2}^{2}}=\Lambda+ k_{4}\rho,\label{fe4}\\
&&\frac{2\ddot a_{2}}{a_{2}}+\frac{\dot
a_{2}^{2}}{a_{2}^{2}}+\frac{1}{a_{2}^{2}}=\Lambda-k_{4}
\omega_{r}\rho -k_{4}p_{0r},\label{fe5}\\
&&\frac{\ddot a_{1}}{a_{1}}+\frac{\ddot a_{2}}{a_{2}}+\frac{\dot
a_{1}\dot a_{2}}{a_{1}a_{2}}=\Lambda-k_{4}\omega_{t}\rho -k_{4}
p_{0t}\label{fe6}.
\end{eqnarray}
The static state corresponding to the above system of nonlinear differential
equations is defined as   $\dot a_1(t)=\dot a_2(t)=\dot \rho(t)=0$. Then,
considering the identifications $a_{1}=a_{01}$, $a_{2}=a_{02}$ and $\rho=\rho_{0}$
for the static state, through the field equations (\ref{fe4})-(\ref{fe6}),
we obtain \begin{eqnarray}
&&\frac{1}{a_{02}^{2}}=\Lambda+ k_{4}\rho_{0},\label{e8}\\
&&\label{e9}
\frac{1}{a_{02}^{2}}=\Lambda-k_{4} \omega_{r}\rho_{0} -k_{4}p_{0r},\\
&&\label{e10}
0=\Lambda-k_{4} \omega_{t}\rho_{0} -k_{4} p_{0t}.
\end{eqnarray}
From (\ref{e10}), the sign of $\Lambda$ depends on the sign of $\omega_t$ and $p_{0t}$. Using equations (\ref{e8}) and (\ref{e9}), we have
\begin{equation}\label{e12}
p_{0r}=-(1+\omega_{r})\rho_{0}=\omega_{eff}\rho_{0}.
\end{equation}
From (\ref{e10}), one gets the
following relation for $p_{0t}$
\begin{equation}\label{e15}
p_{0t}=-\omega_{t}\rho_{0}+\frac{\Lambda}{k_{4}}.
\end{equation}
The equations (\ref{e12}) and (\ref{e15}) represent the constant
radial and lateral effective equations of state of the fluid
supporting the static state.

Now, to study the stability of the static state given in Eqs. (\ref{e8})-(\ref{e10}), we apply the
scalar perturbations to the dynamical quantities of the system (\ref{fe4})-(\ref{fe6}), i.e. the scale factors $a_1(t)$ and $a_2(t)$ and
 the energy density $\rho(t),$ as
\begin{eqnarray}\label{e16}
&&a_{1}(t)\longrightarrow a_{01}\left(1+\delta a_{1}(t)\right),\nonumber\\
&&a_{2}(t)\longrightarrow a_{02}\left(1+\delta a_{2}(t)\right),\nonumber\\
&&\rho(t)\longrightarrow \rho_{0}\left(1+\delta \rho(t)\right).
\end{eqnarray}
Hence, the perturbed  field equations up to the first order take the following
form
\begin{eqnarray}\label{e17}
&&\frac{1}{a^{2}_{02}}-\frac{2\delta
a_2}{a^{2}_{02}}=\Lambda+k_{4}\rho_{0}+ k_{4}\rho_{0}\delta
\rho,\\
&&\label{e18} 2\delta \ddot a_{2}+\frac{1}{a^{2}_{02}}-\frac{2\delta
a_2}{a^{2}_{02}}= \Lambda -k_{4}\omega_{r}\rho_{0}\delta \rho-
k_{4}\omega_{r}\rho_{0}-k_{4}p_{0r},\\
&&\label{e19}
\delta \ddot a_{1}+\delta \ddot a_{2}= \Lambda -k_{4}\omega_{t}\rho_{0}\delta \rho- k_{4}\omega_{t}\rho_{0}-k_{4}p_{0t}.
\end{eqnarray}
Using the constraints given by equations (\ref{e8})-(\ref{e10}),
one can reduce these perturbation equations to
\begin{eqnarray}\label{e20}
&&\frac{2\delta a_2}{a^{2}_{02}}=- k_{4}\rho_{0}\delta \rho,\\
&&\label{e21}
2\delta \ddot a_{2}-\frac{2\delta a_2}{a^{2}_{02}}= -k
_{4}\omega_{r}\rho_{0}\delta \rho,\\
&&\label{e22}
\delta \ddot a_{1}+\delta \ddot a_{2}= -k
_{4}\omega_{t}\rho_{0}\delta \rho.
\end{eqnarray}
Substituting  (\ref{e20})  in  (\ref{e21}) leads to
\begin{equation}\label{e24}
\delta \ddot a_{2}+\gamma^{2}\delta a_2 =0,
\end{equation}
where
\begin{equation}\label{ee28}
\gamma^{2}=-\frac{1+\omega_{r}}{a_{02}^{2}}=\frac{\omega_{eff}}{a_{02}^{2}}.
\end{equation}
Hence, one realizes that the oscillating modes for $\delta a_2$
\begin{eqnarray}\label{e28}
\delta a_{2}=C_{1}e^{i\gamma t}+C_{2}e^{-i\gamma t},
\end{eqnarray}
requires
the constraint\begin{equation}\label{e25}
\omega_{r}<-1.
\end{equation}
Similarly, using (\ref{e20}), (\ref{e21}) and (\ref{e22}), we obtain
\begin{equation}\label{e29}
2\delta \ddot
a_{1}=k_{4}(1+\omega_{r}-2\omega_{t})\rho_{0}\delta\rho.
\end{equation}
Combining  (\ref{e20}) and (\ref{e21}) gives
\begin{equation}\label{e31}
\delta\rho=\frac{-2\delta \ddot a_{2}}{k_{4}(1+\omega_{r})\rho_{0}},
\end{equation}
where substituting in (\ref{e29}) yields\begin{equation}\label{e32}
\delta \ddot
a_{1}=\left(\frac{2\omega_{t}}{1+\omega_{r}}-1\right)\delta \ddot
a_{2},
\end{equation}
which  represents the relation between the radial and lateral perturbations.
Twice integration of this equation gives
\begin{eqnarray}\label{e33}
\delta a_{1}=\left(\frac{2\omega_{t}}{1+\omega_{r}}-1\right)\delta
a_{2}+ C_{3}t+C_{4}
\end{eqnarray}
where $C_3$ and $C_4$ are integration constants. Regarding
(\ref{e28}), the above equation implies  that the stable oscillatory
modes in $\delta a_{1}$ is subjected to the condition $C_{3}=C_4=0$. In this case, the
amplitude of the perturbations along the radial direction depends
explicitly on the nature of the fluid, i.e on the equations of state
parameters. For the particular case $\omega_t=1+\omega_r$, the
perturbation amplitude on the radial and lateral directions are the
same.

In the following,  we elaborate on two specific forms of the fluid (\ref{eos1}) and discuss on the stability
of KSSS.
\\
\\
{\bf 2.1.1.1 Perfect fluid}
\\

Considering a perfect fluid form for the supporting matter fields by setting
$\omega_{r}=\omega_{t}=\omega$ and $p_{0r}=p_{0t}=0$,  equations (\ref{e8}) to (\ref{e10}) for the
static state reduce to
 \begin{eqnarray}
&&\label{e34}
\frac{1}{a_{02}^{2}}=\Lambda+ k_{4}\rho_{0},\\
&&\label{e35}
\frac{1}{a_{02}^{2}}=\Lambda-k_{4}\omega\rho_{0},\\
&&\label{e36}
0=\Lambda-k_{4} \omega\rho_{0}.
\end{eqnarray}
Substituting $p_{0r}=0$ into Eq.(\ref{e12}), one gets either $w_r=-1$ or
$\rho_0=0$. The same result can be observed by comparing
the equations (29) and (30).   If $\rho_0=0$,   Eqs.(\ref{e34}) and (\ref{e35}) will be source free (because of (\ref{e36})) and cannot be held anymore for finite $a_0$.
If $\omega=\omega_r=\omega_t=-1$ for a perfect fluid, Eq.(\ref{e36})  (which
demands
$\Lambda<0$)  is inconsistent
with Eqs.(\ref{e34}) and (\ref{e35}).  In either way, one observes  that having a finite size static
sate for the KS spacetime is impossible.
Then, a perfect fluid cannot support a nonsingular KSSS. Hence, for the realization of a stable nonsingular
KSSS, the modification of the perfect  fluid from is inevitable.
\\
\\
{\bf 2.1.1.2 Anisotropic fluid}
\\

Here, we consider two possible minimal modifications of the perfect
fluid form leading to  anisotropic fluids: $(1)$ a fluid having
different equations of state
parameters on the radial and
lateral directions, and $(2)$ a fluid modifying the perfect fluid
form with distinct constant pressures on the radial and lateral
directions.
\\
\\
{\bf Type 1}:
\\
In this case, considering two different equations of state
parameters for the radial and
lateral directions we have
\begin{equation}\label{e38}
p_{r}=\omega_{r}\rho,~~~~p_{t}=\omega_{t}\rho.
\end{equation}
Then, equations (\ref{e8}) to (\ref{e10}) governing the static state
reduce to\begin{eqnarray} &&\label{e39}
\frac{1}{a_{02}^{2}}=\Lambda+ k_{4}\rho_{0},\\
&&\label{e40}
\frac{1}{a_{02}^{2}}=\Lambda-k_{4}\omega_{r}\rho_{0},\\
&&\label{e41} 0=\Lambda-k_{4}\omega_{t}\rho_{0}.
\end{eqnarray}
Satisfaction of (\ref{e41}) demands that $\Lambda$ and $\omega_t$ have the same sign. Comparing equations (\ref{e39}) and (\ref{e40}) we obtains the constraint
$\omega_{r}=-1$ on the equation of state in the radial direction to have
a nonsingular static state.   On the other hand, we have seen that the
stability of KSSS demands $\omega_{r}<-1$,  i.e. equation (\ref{e25}). Therefore,
one can conclude that the anisotropic fluid from defined in (\ref{e38}) can support a
static state but this static state is not stable versus the scalar perturbations (\ref{e16}). \\
\\
{\bf Type 2}:
\\
We consider a fluid modifying
the perfect fluid form by distinct  constant pressures
on the radial and lateral  directions.
For this aim, we introduce  the constant pressure terms
$p_{0t}$ and $p_{0r}$ to the equations of state of the fluid with
$\omega_{r}=\omega_{t}=\omega$. In this case, the  equations
in (\ref{eos1}) become
\begin{equation}\label{e43}
p_{r}=\omega\rho +p_{0r},~~~~p_{t}=\omega\rho +p_{0t}.
\end{equation}
Then  equations  (\ref{e12}) and (\ref{e15}) for the static state
reduce to\begin{eqnarray}\label{e44}
&&p_{0r}=-(1+\omega)\rho_{0},\nonumber\\
&&\label{e45}
p_{0t}=-\omega\rho_{0}+\frac{\Lambda}{k_{4}}.
\end{eqnarray}
Combining them gives the following relation between the constant pressures
 $p_{0r}$ and $p_{0t}$
\begin{equation}\label{e46}
p_{0t}-p_{0r}=\rho_{0}+\frac{\Lambda}{k_{4}}.
\end{equation}
In the case $\rho_{0}+\frac{\Lambda}{k_{4}}=0$, we face an inconsistency in the static state given by equations
(\ref{e8}) to (\ref{e10}). Thus, to have a static state we consider
$\rho_{0}+\frac{\Lambda}{k_{4}}\neq0$. In this case, $\gamma^2$ given by equation (\ref{ee28}) becomes
\begin{eqnarray}\label{eee28}
\gamma^{2}=-\frac{1+\omega}{a_{02}^{2}}.
\end{eqnarray}
Therefore, the positivity condition on $\gamma^2$ to have a stable nonsingular   static state
demands
 \begin{equation}\label{omega}
 \omega<-1,
 \end{equation}
 which means that the fluid supporting the geometry lies in the phantom range.

For the stability of this case, the oscillatory modes of
 $\delta a_{2}$ is given by equation  (\ref{e28}) with $\gamma$ in (\ref{eee28}). The dynamics of  $\delta a_{1}$ from equation (\ref{e33}) reads as
\begin{equation}\label{e47}
\delta a_{1}=\left(\frac{\omega -1}{\omega+1}\right)\delta a_{2}+
C_{3}t+C_{4},
\end{equation}
where for $C_{3}=C_4=0$, the oscillatory modes are possible and hence the
static state will be stable.
Regarding the stability requirement (\ref{omega}), one notes that although the perturbation
in both directions have the same sign but the amplitude of the perturbations on the
radial direction is always greater than the perturbations on the lateral direction.

\subsubsection{Energy-momentum source with generalized Chaplygin gas
equation of state } Here, we consider a generalization of the
Chaplygin gas type fluid \cite{chap1, chap2} for the energy-momentum
source possessing\begin{equation}\label{eos2}
p_{r}=-\frac{\alpha_r}{\rho^n},~~~~p_{t}=-\frac{\alpha_t}{\rho^m}.
\end{equation}
 where $0\leq m,n\leq 1$, $\alpha_r$ and $\alpha_t$ are positive constants. If $\alpha_r=\alpha_t$
 and $m=n=1$,
 the energy-momentum source reduces to the generalized  Chaplygin gas
 in \cite{chap2}. The equations
governing a static state have the form
\begin{eqnarray}
&&\frac{1}{a_{02}^{2}}=\Lambda+ k_{4}\rho_{0},\label{c7}\\
&&\label{c8}
\frac{1}{a_{02}^{2}}=\Lambda+ k_{4}\frac{\alpha_r}{\rho_0^n},\\
&&\label{c9} 0=\Lambda+ k_{4} \frac{\alpha_t}{\rho_0^m}.
\end{eqnarray}
Satisfaction of (\ref{c9}) demands  $\Lambda<0$ for $\alpha_t>0$. Then, using (\ref{e16}),  the perturbed
field equations around the static state given by equations (\ref{c7})-(\ref{c9}) reduce to
\begin{eqnarray}\label{c13}
&&\frac{2\delta a_2}{a^{2}_{02}}= -k_{4}\rho_{0}\delta
\rho,\\
&&\label{c14} 2\delta \ddot a_{2}-\frac{2\delta
a_2}{a^{2}_{02}}= -nk_4\frac{\alpha_r}{\rho_0^n}\delta\rho,\\
&&\label{c15} \delta \ddot a_{1}+\delta \ddot
a_{2}=-mk_{4}\frac{\alpha_t}{\rho_0^m}\delta\rho.
\end{eqnarray}
Combining (\ref{c13}) and (\ref{c14}), we obtain\begin{equation}\label{c16}
\delta \ddot a_{2}+\gamma^{2}\delta a_2 =0,
\end{equation}
where
\begin{equation}
\gamma^{2}=-\frac{n\alpha_r \rho_0^{-(n+1)}+1}{a_{02}^{2}}.
\end{equation}
Here, one observes that since $\alpha_r>0 $, then $\gamma^{2}<0$ and consequently
there
are no oscillatory modes for $\delta a_2$ and $\delta a_1$.

We recapitulate our analysis in this section as follows. \textit{It is proved that in the context of
Einstein's GR in four dimensions:
$(i)$ A perfect
fluid source cannot support a finite size static KS geometry, $(ii)$
An anisotropic fluid with linear equations of state  $p_{r}=\omega_{r}\rho$ and $p_{t}=\omega_{t}\rho~$
supports a finite size static KS geometry but it is not stable
against the scalar perturbations,  $(iii)$ A modification in the linear equation
of state of the perfect
fluid form as $p_{r}=\omega\rho +p_{0r}$ and $p_{t}=\omega\rho +p_{0t}$ can
support a stable nonsingular KS type static state, and $(iv)$
 A generalized Chaplygin gas fluid having $p_{r}=-\frac{\alpha_r}{\rho^n},~p_{t}=-\frac{\alpha_t}{\rho^m}$ cannot support a stable nonsingular KS geometry in four spacetime
dimensions.}

In the following section, since anisotropic models are exact
solutions to the string theory, we investigate the existence and
stability conditions for a KSSS in a five dimensional gravity
theory, and show how the extra dimensional
geometric modifications  affect the results in four dimensions.%%%%%%%%%%%%%%%%%%%%%%%%%%%%%%%%%%%%%%%%%%%%%%%%%%%%%%%%%%%%%%%%%%%%%%%%%%%%%
\section{KS geometry on the brane}
We consider the five dimensional
action  \cite{Shiro}
\begin{eqnarray}\label{ac}
S=\int d^5 x\sqrt{-\mathcal{G}}\left(\frac{1}{2k_5}\mathcal{R}-\Lambda_5\right)
+\int_{\xi=0}
d^4 x\sqrt{-g}\left(\frac{1}{k_5}K^{\pm}-\lambda+L_{matter}\right),
\end{eqnarray}
where $k_5=8\pi G_5$, $\mathcal{G}$, $g$, $\mathcal{R}$, $\Lambda_5$, $\lambda$,
and $L_{matter}$  are the 5-dimensional gravitational coupling constant, trace of the bulk space's metric $\mathcal{G}_{AB}$,
trace of the brane's metric $g_{\mu\nu}$, bulk space's Ricci scalar, bulk   and  brane vacuum energy, and the Lagrangian of the confined matter fields
to brane, respectively. Also,
$x^\mu$ with $\mu=0,...,3$ represents the coordinates of the brane
while $\xi$ represents the single extra dimension orthogonal to the brane, and $K^\pm$ is the extrinsic
curvature on either sides of the brane.

Variation of the action (\ref{ac}) with respect to the bulk metric $\mathcal{G}_{AB}$
gives the $5D$ Einstein field equations
\begin{eqnarray}
&&^{(5)}G_{ab}=k_5 {}^{(5)}T_{ab},\nonumber\\
&&{}^{(5)}T_{ab}=-\Lambda_5 \mathcal{G}_{ab}+
\delta(\xi)\left(-\lambda \mathcal{G}_{ab}+T_{ab}^{matter}\right),
\end{eqnarray}
where $a, b=0, ..., 4$. In order to keep the generality of our analysis,
similar to the four dimensional case,  we consider the  energy-momentum tensor $T^{matter}_{\mu\nu}$  corresponding to $L_{matter}$  on the brane as $T^\mu_\nu=diag\left(-\rho, \,p_r, \,p_t, \,p_t\right)$  \cite{Ram}. Considering the bulk space metric in the form
\begin{equation}
 ds^2=\left(N_a N_b+g_{ab}\right)dx^a dx^b,
 \end{equation}
  where
$N^a $ represents  the unit normal vector to the hypersurface $\xi=constant$  and $g_{ab}$
is the induced metric on this hypersurface, the induced  field
equations
on the brane take the following form \cite{Sasaki}
\begin{equation}
\label{qw}
G_{\mu\nu}=-\Lambda g_{\mu\nu}+k_4 T_{\mu\nu}+k_5^2 S_{\mu\nu}-E_{\mu\nu},
\end{equation}
in which\begin{eqnarray}
S_{\mu\nu}=\frac{1}{12}T T_{\mu\nu}-\frac{1}{4}T_\mu^\alpha T_{\alpha\nu}
+\frac{1}{24}g_{\mu\nu}(3T^{\alpha\beta}T_{\alpha\beta}-T^2).\nonumber
\end{eqnarray}
Here $\Lambda=k_5(\Lambda_5+k_5\lambda^2/6)$ is the effective
cosmological constant on the brane, and $E_{ab}=C_{abcd}N^a N^b$
where $C_{abcd}$ represents the 5-dimensional Weyl tensor of the
bulk space.

Considering the brane cosmology with zero Weyl tensor and
a generic energy-momentum tensor possessing the form we mentioned above, the Einstein field equations for the metric (\ref{11}) on the brane read
as \begin{eqnarray}
&&\label{1a} \frac{\dot a_{2}^{2}}{a_{2}^{2}}+\frac{2\dot a_{1}\dot
a_{2}}{a_{1}a_{2}}+\frac{1}{a_{2}^{2}}=\Lambda+ k_{4}\rho +
\frac{1}{12}k_{5}^{2}(\rho^{2}-p_{r}^{2}-p_{t}^{2}+2p_{r}p_{t}),\\
&&\label{1b} \frac{2\ddot a_{2}}{a_{2}}+\frac{\dot
a_{2}^{2}}{a_{2}^{2}}+\frac{1}{a_{2}^{2}}=\Lambda- k_{4}p_{r}-
\frac{1}{12}k_{5}^{2}(\rho^{2}-p_{r}^{2}+p_{t}^{2}+2\rho p_{t}),\\
&&\label{1c} \frac{\ddot a_{1}}{a_{1}}+\frac{\ddot
a_{2}}{a_{2}}+\frac{\dot a_{1}\dot a_{2}}{a_{1}a_{2}}=\Lambda-
k_{4}p_{t}- \frac{1}{12}k_{5}^{2}(\rho^{2}+p_{r}^{2}+\rho p_{r}+\rho
p_{t}-p_{r}p_{t}).
\end{eqnarray}
\subsection{KSSS and stability analysis on the brane}
In the following, we investigate the existence and stability of KS type static state on the 4D brane  considering the field equations (\ref{1a})-(\ref{1c}). For the sake of generality of the analysis, we study two generic kinds of energy-momentum sources: $(i)$ the sources possessing linear equation of
state, and $(ii)$  the sources with generalized
Chaplygin gas type equation of state.
\subsubsection{Energy-momentum source with linear equations of state}
We consider a general equation of state with the form given in
(\ref{eos1}), and  introduce the parameters
\begin{eqnarray}\label{AtoI}
&&A=\frac{1}{12}k_{5}^{2}(1-\omega_{r}^{2}-\omega_{t}^{2}+2\omega_{r}\omega_{t}),\nonumber\\
&&B=k_{4}+\frac{1}{6}k_{5}^{2}\left(\omega_{r}(p_{0t}-p_{0r})+\omega_{t}(p_{0r}-p_{0t})\right),\nonumber\\
&&C=\frac{1}{12}k_{5}^{2}(2p_{0r}p_{0t}-p_{0r}^2-p_{0t}^2),\nonumber\\
&&D=\frac{1}{12}k_{5}^{2}(1-\omega_{r}^{2}+\omega_{t}^2+2\omega_{t}),\nonumber\\
&&E=\frac{1}{6}k_{5}^{2}(\omega_{t}p_{0t}-\omega_{r}p_{0r}+p_{0t})+k_{4}\omega_{r},\nonumber\\
&&F=\frac{1}{12}k_{5}^{2}(p_{0r}^2-p_{0t}^2)-k_{4}p_{0r},\nonumber\\
&&G=\frac{1}{12}k_{5}^{2}(1+\omega_{r}^{2}+\omega_{r}+\omega_{t}-\omega_{r}\omega_{t}),\nonumber\\
&&H=k_{4}\omega_{t}+\frac{1}{12}k_{5}^{2}\left(\omega_{r}(2p_{0r}-p_{0t})-\omega_{t}p_{0r}+p_{0r}+p_{0t}\right),\nonumber\\
&&I=\frac{1}{12}k_{5}^{2}(p_{0r}p_{0t}-p_{0r}^2)-k_{4}p_{0t}.
\end{eqnarray}
Hence the field equations (\ref{1a})-(\ref{1c}) reduce to the following forms
\begin{eqnarray}
&&\label{f5} \frac{\dot a_{2}^{2}}{a_{2}^{2}}+\frac{2\dot a_{1}\dot
a_{2}}{a_{1}a_{2}}+\frac{1}{a_{2}^{2}}=\Lambda+A\rho^{2}+B\rho+C,\\
&&\label{f6} \frac{2\ddot a_{2}}{a_{2}}+\frac{\dot
a_{2}^{2}}{a_{2}^{2}}+\frac{1}{a_{2}^{2}}=\Lambda- D\rho^{2} -E\rho
+F,\\
&&\label{f7} \frac{\ddot a_{1}}{a_{1}}+\frac{\ddot
a_{2}}{a_{2}}+\frac{\dot a_{1}\dot a_{2}}{a_{1}a_{2}}=\Lambda-
G\rho^{2} -H\rho +I.
\end{eqnarray}
For a static state defined by  $\dot a_{1}=\dot a_{2}=\ddot a_{1}=\ddot a_{2}=0$,
let $a_{1}=a_{01}$, $a_{2}=a_{02}$ and $\rho=\rho_{0}$. Then, the field equations (\ref{f5})-(\ref{f7}) give
\begin{eqnarray}
&&\label{f8}
\frac{1}{a_{02}^{2}}=\Lambda+A\rho_{0}^{2}+B\rho_{0}+C,\\
&&\label{f9} \frac{1}{a_{02}^{2}}=\Lambda- D\rho_{0}^{2} -E\rho_{0}
+F,\\
&&\label{f10} 0=\Lambda- G\rho_{0}^{2} -H\rho_{0}+I.
\end{eqnarray}
The Using equations (\ref{f8}) and (\ref{f9}), we have the
 constraint equation  for
$\rho_{0}$
\begin{equation}\label{rho}
(A+D)\rho_{0}^2+(B+E)\rho_{0}+C-F=0,
\end{equation}

\begin{figure}[ht] 
\centering
\includegraphics[scale=0.55]{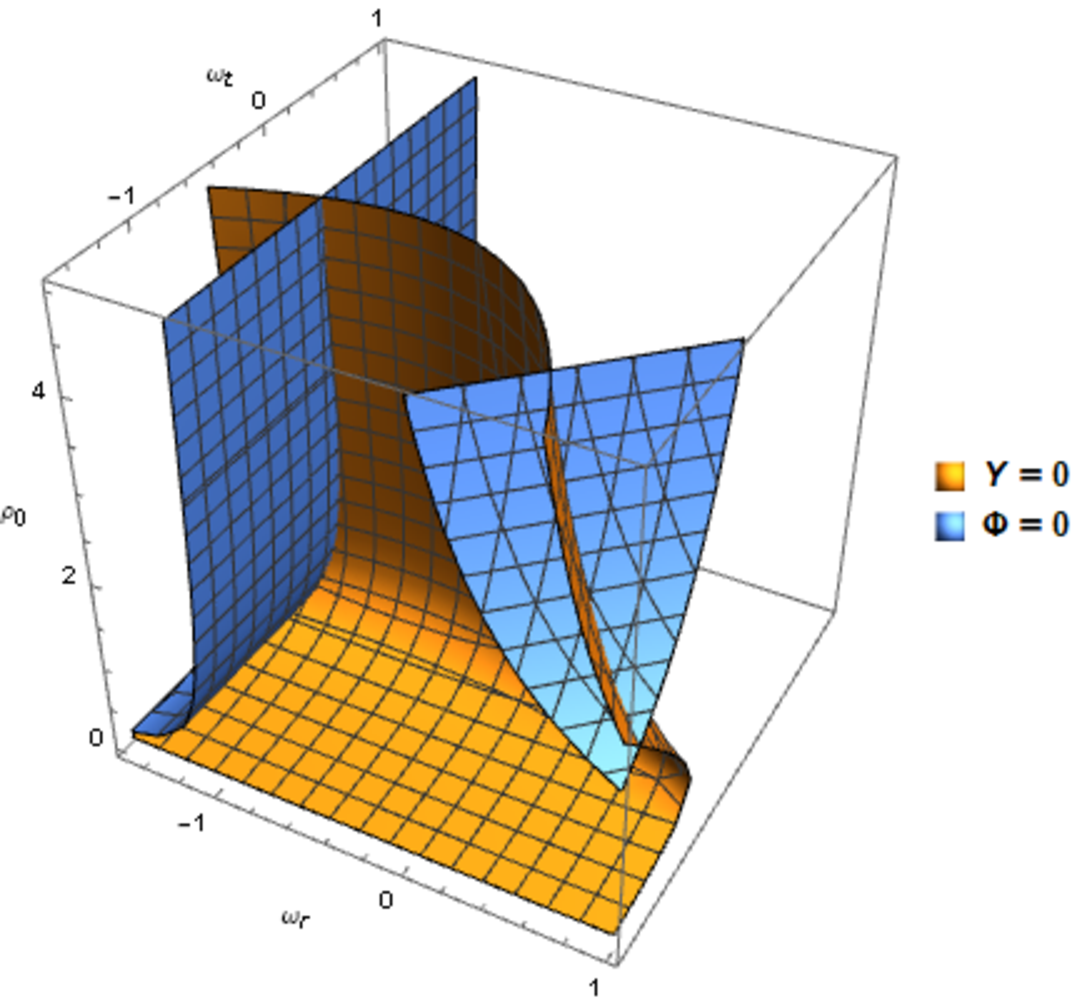}
\includegraphics[scale=0.55]{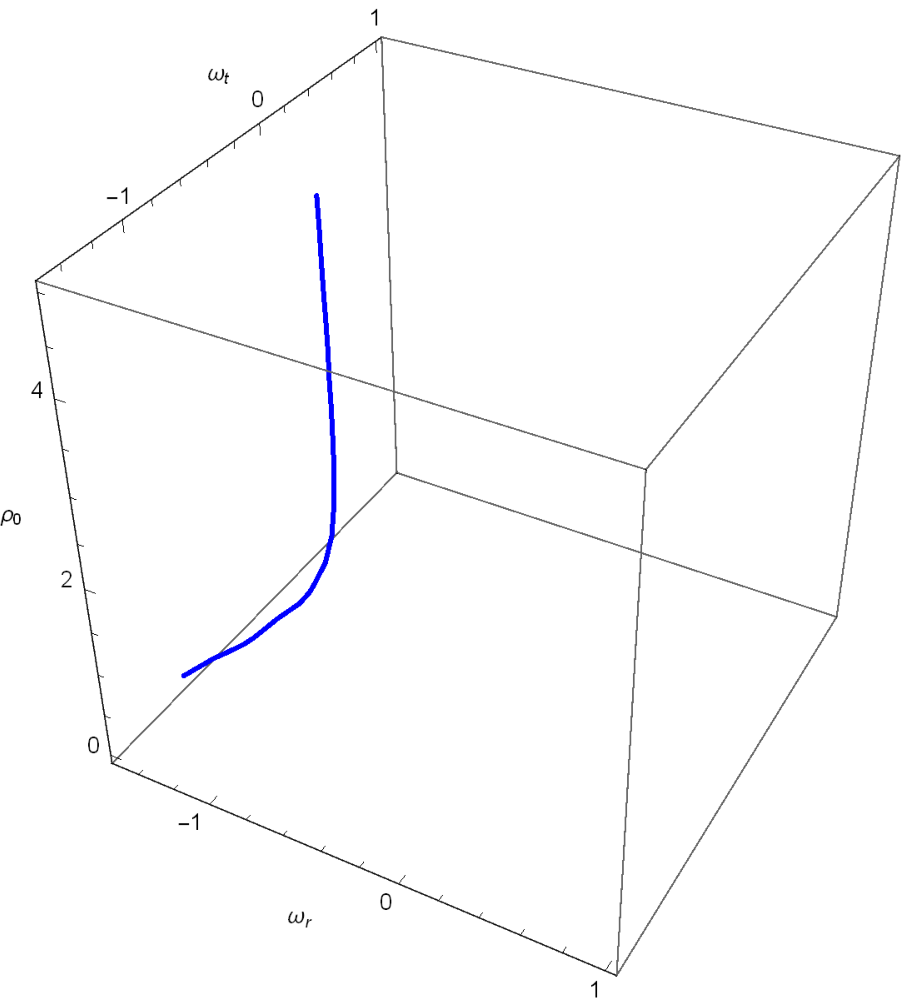}
\includegraphics[scale=0.55]{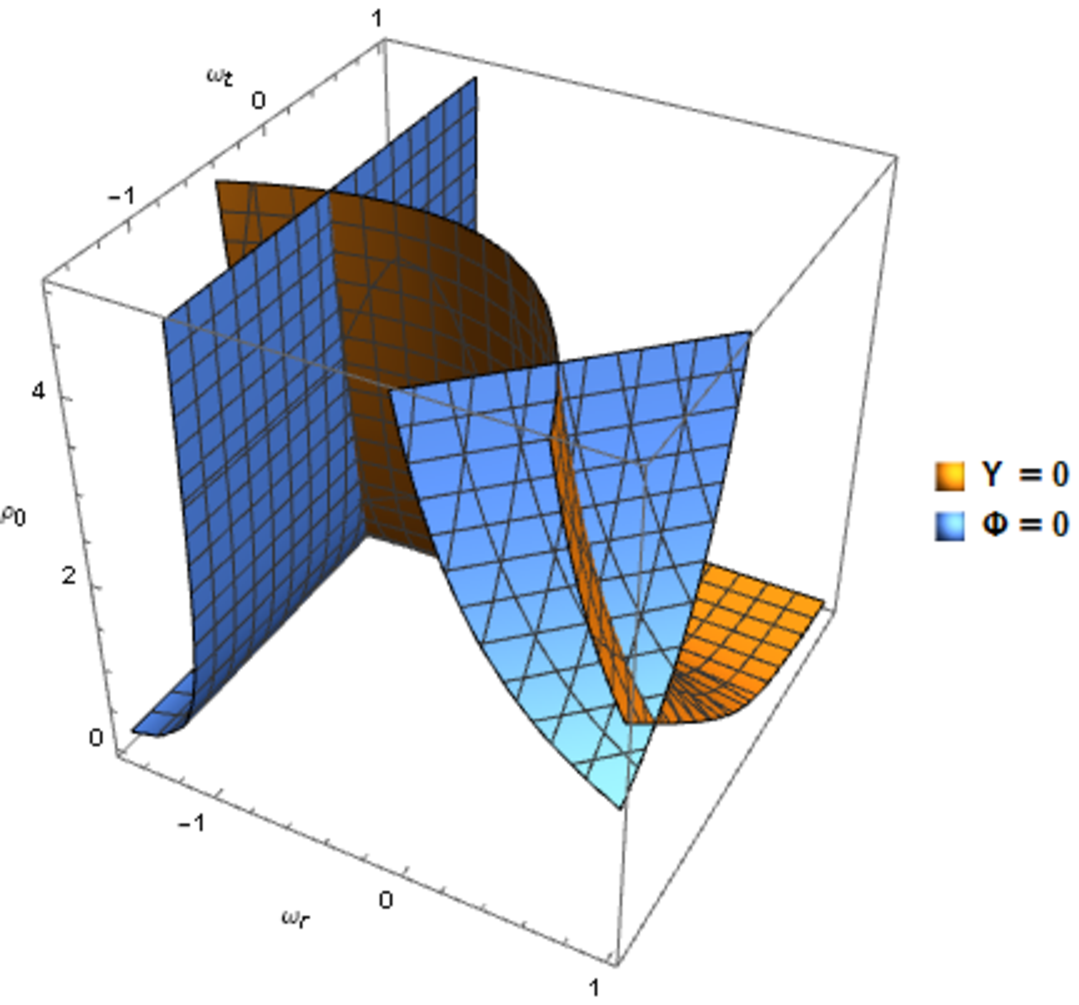}
\includegraphics[scale=0.55]{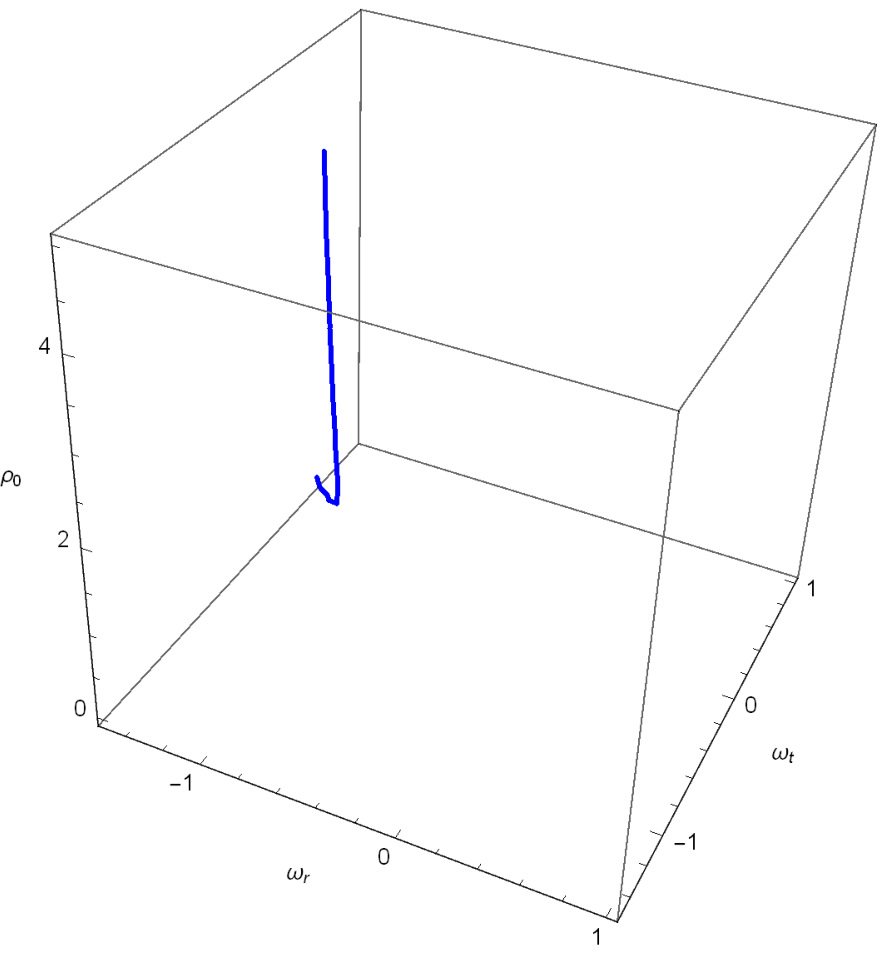}
\caption{\label{fig1} The left plots
show the intersecting surfaces corresponding to constraints
(\ref{f10}) and (\ref{rho}) for having a static state. $Y=0$
represents equation (\ref{f10}) and $\Phi=0$ represents equation
(\ref{rho}). The right plots show the intersection of surfaces for
the existence conditions in the solution space of stability conditions, i.e equations
(\ref{f25}) and (\ref{beta2}).
The curves in the right plots satisfy all the conditions
(\ref{f10}),(\ref{rho}),(\ref{f25}) and (\ref{beta2}) to have a
totally stable static state. We used the typical values of $p_{0r}=0.1$ and
$p_{0t}=0.2$ for the first line and $p_{0r}=0.1$ and $p_{0t}=-0.1$
for the second line. We also assumed $k_{4}=1$, $k_{5}=1.5$,
$\Lambda=10^{-5}$.}
\end{figure}

Now, to study the stability of static state defined in (\ref{f8})-(\ref{f10}), we consider the scalar
perturbations in the form of equations (\ref{e16}). Keeping up to the
first order perturbation terms, equations (\ref{f5}) to (\ref{f7})
give
\begin{eqnarray}
&&\label{f17} \frac{1}{a^{2}_{02}}-\frac{2\delta
a_2}{a^{2}_{02}}=\Lambda+A\rho_{0}^{2}(1+2\delta\rho)+B\rho_{0}(1+\delta\rho)+C,\\
&&\label{f18} 2\delta \ddot a_{2}+\frac{1}{a^{2}_{02}}-\frac{2\delta
a_2}{a^{2}_{02}}= \Lambda- D\rho_{0}^{2}(1+2\delta\rho)
-E\rho_{0}(1+\delta\rho)+F,\\
&&\label{f19} \delta \ddot a_{1}+\delta \ddot a_{2}= \Lambda-
G\rho_{0}^{2}(1+2\delta\rho)-H\rho_{0}(1+\delta\rho)+I.
\end{eqnarray}
Using the constraints given by equations (\ref{f8}) to (\ref{f10}),
one can reduce above equations  to
\begin{eqnarray}
&&\label{f20} -\frac{2\delta
a_2}{a^{2}_{02}}=(2A\rho_{0}^{2}+B\rho_{0})\delta\rho,\\
&&\label{f21} 2\delta \ddot a_{2}-\frac{2\delta
a_2}{a^{2}_{02}}=(-2D\rho_{0}^{2}-E\rho_{0})\delta\rho,\\
&&\label{f22} \delta \ddot a_{1}+\delta \ddot
a_{2}=(-2G\rho_{0}^{2}-H\rho_{0})\delta\rho.
\end{eqnarray}
From equation (\ref{f20}), we have
\begin{equation}\label{m20}
\delta\rho=-\frac{2}{a^{2}_{02}(2A\rho_{0}^{2}+B\rho_{0})}\delta
a_2,
\end{equation}
where substituting in equation (\ref{f21}) leads to
\begin{equation}\label{f24}
\delta \ddot a_{2}+\alpha^2\delta a_2 =0,
\end{equation}
where
\begin{equation}\label{alpha}
\alpha^{2}=-\frac{1}{a_{02}^{2}}\left(1+\frac{2D\rho_{0}+E}{2A\rho_{0}+B}\right).
\end{equation}
Hence, the oscillating modes for $\delta a_2$ requires \begin{equation}\label{f25}
1+\frac{2D\rho_{0}+E}{2A\rho_{0}+B}<0.
\end{equation}
The solution to equation (\ref{f24}) with the condition (\ref{f25})
is
\begin{eqnarray}\label{f28}
\delta a_{2}=C_{1}e^{i\alpha t}+C_{2}e^{-i\alpha t},
\end{eqnarray}
which shows the oscillatory behavior of $\delta a_{2}$. \\
On the other
hand, combining equations (\ref{f20}), (\ref{f21}) and (\ref{f22}) gives
\begin{equation}\label{f29}
2\delta\ddot
a_{1}=((2A+2D-4G)\rho_{0}^{2}+(B+E-2H)\rho_{0})\delta\rho,
\end{equation}
and subtracting (\ref{f21}) from (\ref{f20}) leads to
\begin{equation}\label{f30}
2\delta \ddot a_{2}=-((2A+2D)\rho_{0}^{2}+(B+E)\rho_{0})\delta\rho.
\end{equation}
Hence, using (\ref{f30}) and
(\ref{f29}) we obtain\begin{equation}\label{f32}
\delta \ddot a_{1}=\beta\delta \ddot a_{2},
\end{equation}
where $\beta$ is defined as
\begin{equation}\label{beta}
\beta=\frac{4G\rho_{0}+2H}{(2A+2D)\rho_{0}+B+H}-1.
\end{equation}
Twice integration of equation (\ref{f32}) gives
\begin{eqnarray}\label{f33}
\delta a_{1}=\beta\delta a_{2}+ C_{3}t+C_{4}.
\end{eqnarray}
This shows that $\delta a_{1}$ can also have an stable oscillatory
mode if $C_{3}=C_4=0$, and
\begin{equation}\label{beta2}
2(A+D)\rho_{0}+B+H\neq0.
\end{equation}
Hence, both the conditions in (\ref{f25}) and (\ref{beta2}) should
be satisfied to have a stable static state. This result holds for a
fluid with the general form of equation of state (\ref{eos1}). In
Figures \ref{fig1}, the existence and stability conditions (i.e.
equations (\ref{f10}), (\ref{rho}), (\ref{f25}) and (\ref{beta2}))
for a static state are plotted for some typical values of the
parameters. The presence of some $\rho_0>0$ ranges represents the
satisfaction of the constraints that means a stable static state
exists for the given values of the parameters. The intersection of
$Y=0$ and $\Phi=0$ lies in the range $\omega_{r}<-1$ that guarantees
a stable static state. When $p_{0r}=0.1$ and $p_{0t}=0.2$,
$\omega_{t}$ should be also negative, but for $p_{0r}=0.1$,
$p_{0t}=-0.1$, and small amounts of $\rho_0>0$ , the constraints for
having a stable static state can be satisfied for a positive value
of $\omega_{t}$. It is also clear from figure \ref{fig1} that, in
this setup, changing the value of $p_{0t}$ did not affect the result
of equation (\ref{rho}), but have a significant effect on equation
(\ref{f10}) for small values of $\rho_0>0$.

In the following, we consider some special types of fluids and discuss under
what conditions these fluids can support a nonsingular KSSS on the
brane.
\\
\\
{\bf 3.1.1.1 Perfect fluid}

In this case, we consider perfect fluid which means
$\omega_{r}=\omega_{t}=\omega$ and $p_{0r}=p_{0t}=0$. Then, the
coefficients A to I defined in equations (\ref{AtoI}) reduce to
\begin{eqnarray}\label{AtoI2}
&&A=\frac{1}{12}k_{5}^{2},\nonumber\\
&&B=k_{4},\nonumber\\
&&D=G=\frac{1}{12}k_{5}^{2}(1+2\omega),\nonumber\\
&&E=H=k_{4}\omega,\nonumber\\
&&C=F=I=0.
\end{eqnarray}
For the the  static state given by equations (\ref{f8}) to
(\ref{f10}), one obtains
\begin{eqnarray}
&&\label{ssp1}
\frac{1}{a_{02}^{2}}=\Lambda+\frac{1}{12}k_{5}^{2}\rho_{0}^{2}+k_{4}\rho_{0},\\
&&\label{ssp2}
\frac{1}{a_{02}^{2}}=\Lambda-\frac{1}{12}k_{5}^{2}(1+2\omega)\rho_{0}^{2}-k_{4}\omega\rho_{0},\\
&&\label{ssp3} 0=\Lambda-
\frac{1}{12}k_{5}^{2}(1+2\omega)\rho_{0}^{2}-k_{4}\omega\rho_{0}.
\end{eqnarray}
Similar to four dimensions, these equations lead to
\begin{equation}\label{ssp4}
\frac{1}{a_{02}^{2}}=0,
\end{equation}
which means that a perfect fluid cannot support a KSSS even in the
presence of higher dimensional modifications.
\\
\\
{\bf 3.1.1.2 Anisotropic fluid}

\begin{figure}[ht] \centering
\includegraphics[scale=0.55]{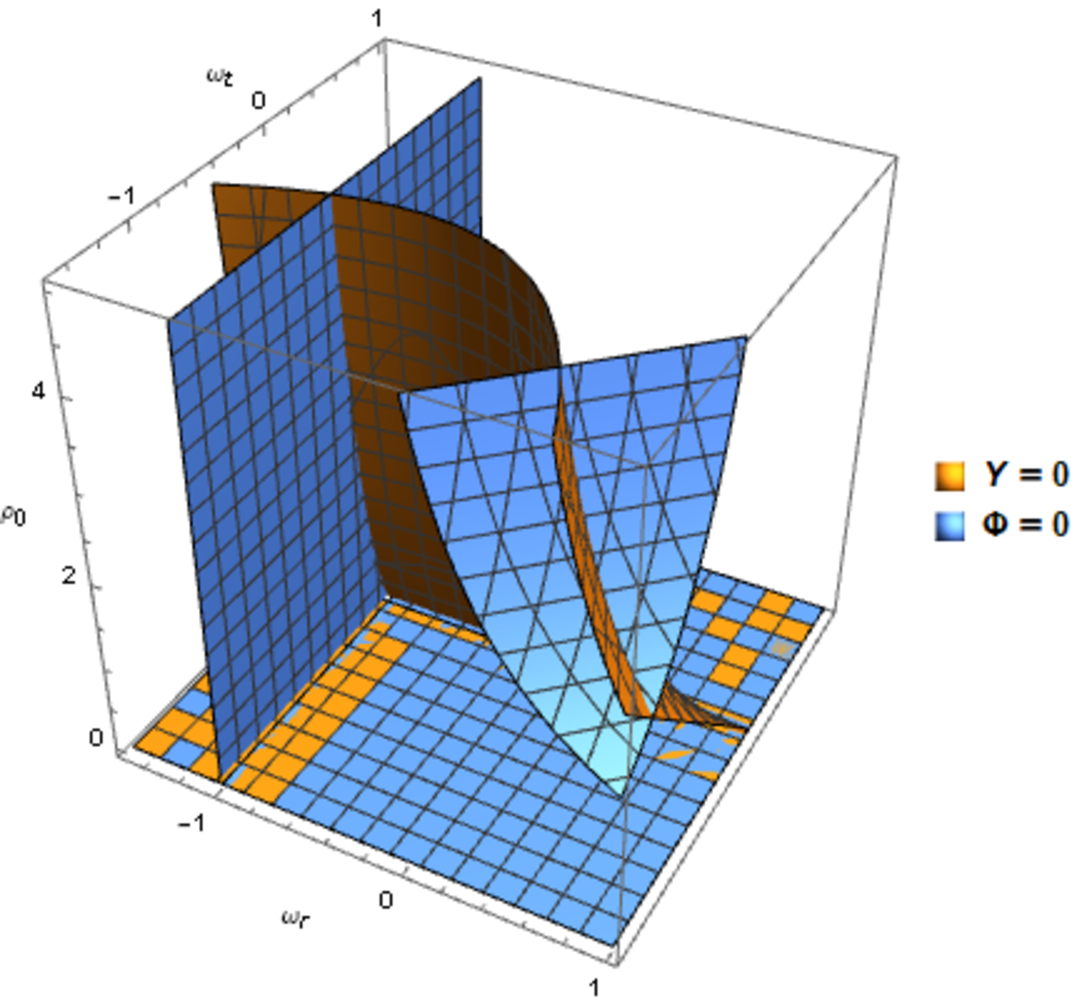}
\includegraphics[scale=0.55]{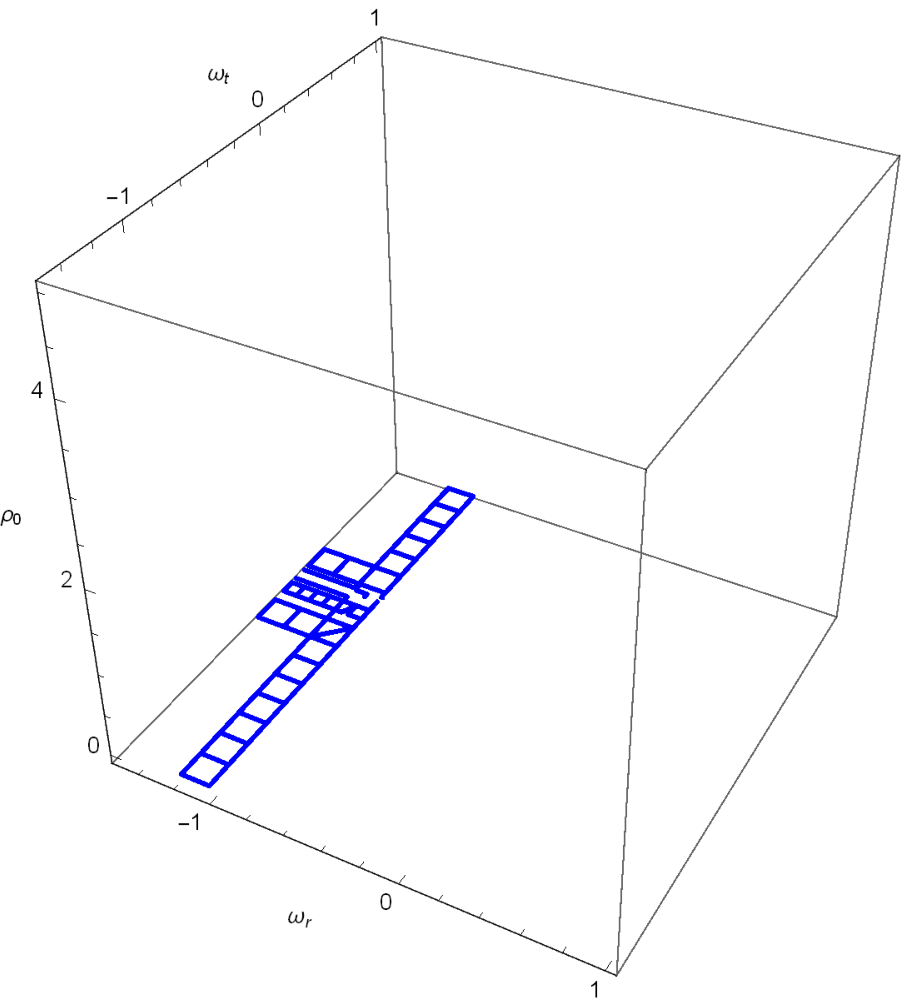}
\caption{\label{fig3}  The first
plot shows the intersecting surfaces corresponding to constraints
(\ref{cons3}) and (\ref{cons4}) for having a static state.  $Y=0$
and $\Phi=0$ represent equations (\ref{cons3}) and (\ref{cons4}),
respectively. The second plot shows the intersection of surfaces for
the existence conditions in the solution space of stability conditions, i.e equations
(\ref{cond}) and (\ref{beta3}).
The value of the parameters in blue in the second plot satisfy all the conditions
(\ref{cons3}),(\ref{cons4}),(\ref{cond}) and (\ref{beta3}) to have a
totally stable static state. The used typical values and ranges are:
$k_{4}=1$, $k_{5}=1.5$, $p_{0r}=0$, $p_{0t}=0$, $\Lambda=10^{-5}$,
$-1.5<\omega_{r},~ \omega_{t}<1$, $0.0001<\rho_{0}<5$.}
\end{figure}

\begin{figure}[ht]
\centering
\includegraphics[scale=0.55]{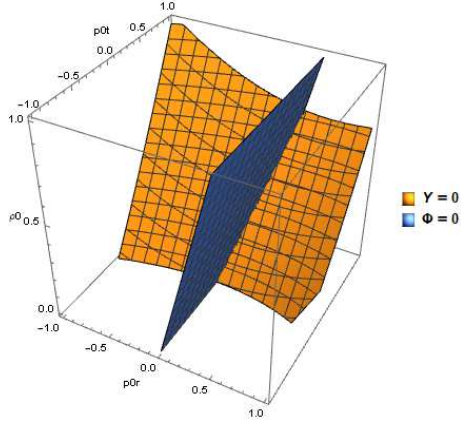}
\caption{\label{general2} The figures show the contours for
constraints (\ref{f44}) and (\ref{f45}) in the solution space of
equations (\ref{cond2}) and (\ref{beta4}). $Y=0$ represents equation
(\ref{f44}) and $\Phi=0$ represents equation (\ref{f45}). The used
typical values and ranges are: $k_{4}=1$, $k_{5}=1.5$,
$\Lambda=10^{-5}$, $-1<p_{0r},~ p_{0t}<1$, and
$\omega=-\frac{3}{2}$. Here, the surfaces $Y=0$ and $\Phi=0$
intersect on a line that means the satisfaction of all the
conditions (\ref{f44}),(\ref{f45}),(\ref{cond2}) and (\ref{beta4}).
}
\end{figure}

In four dimensions, we showed that  a stable static state can exist only
for
an anisotropic fluid of Type 2. Here, we show that in the presence of higher
dimensional modifications to the field equations, both Type 1  and Type 2 fluids
can support a stable nonsingular KSSS.\\
\\
{\bf Type 1}:
\\
In this case we consider equations of state in (\ref{e38}). Hence, the coefficients A to I defined by
equations (\ref{AtoI}) reduce to
\begin{eqnarray}\label{AtoI3}
A&=&\frac{1}{12}k_{5}^{2}(1-\omega_{r}^{2}-\omega_{t}^{2}+2\omega_{r}\omega_{t}),\nonumber\\
B&=&k_{4},\nonumber\\
D&=&\frac{1}{12}k_{5}^{2}(1-\omega_{r}^{2}+\omega_{t}^2+2\omega_{t}),\nonumber\\
E&=&k_{4}\omega_{r},\nonumber\\
G&=&\frac{1}{12}k_{5}^{2}(1+\omega_{r}^{2}+\omega_{r}+\omega_{t}-\omega_{r}\omega_{t}),\nonumber\\
H&=&k_{4}\omega_{t},\nonumber\\
C&=&F=I=0.
\end{eqnarray}
Then the constraints (\ref{f10}) and (\ref{rho}) for the existence
of a static state,  respectively, become
\begin{equation}\label{cons3}
\Lambda
-\left(\frac{1}{12}k_{5}^{2}(1+\omega_{r}^{2}+\omega_{r}+\omega_{t}-\omega_{r}\omega_{t})\right)\rho_{0}^{2}-k_{4}\omega_{t}\rho_{0}=0,
\end{equation}
and
\begin{equation}\label{cons4}
\frac{1}{6}k_{5}^{2}(1-\omega_{r}^{2}+\omega_{t}+\omega_{r}\omega_{t})\rho_{0}^{2}+k_{4}(1+\omega_{r})\rho_{0}=0.
\end{equation}
Also the conditions (\ref{f25}) and (\ref{beta2}) to have an
oscillatory mode, respectively, become
\begin{equation}\label{cond}
\frac{\frac{1}{6}k_{5}^{2}(1-\omega_{r}^{2}+\omega_{t}^2+2\omega_{t})\rho_{0}+k_{4}\omega_{r}}{\frac{1}{6}k_{5}^{2}(1-\omega_{r}^{2}-\omega_{t}^{2}+2\omega_{r}\omega_{t})\rho_{0}+k_{4}}<-1,
\end{equation}
and
\begin{equation}\label{beta3}
\rho_{0}\neq
-\frac{3k_{4}}{k_{5}^{2}\left(1+\frac{\omega_{r}(\omega_{t}-\omega_{r})}{1+\omega_{t}}\right)}.
\end{equation}
Thus, in contrast to four dimensions,  the anisotropic fluid of Type
1 can support a stable nonsingular ESU on the brane under the
conditions (\ref{cons3})-(\ref{beta3}). In Figure \ref{fig3}, these
conditions are plotted for some typical values of the parameters. We
see from the figure that the constraints can be satisfied for
$\rho_0>0$. This means a stable static state do exist for the given
values of the parameters. In this case, it is obvious that to have a
stable static state we need $\omega_{r}<-1$, but $\omega_{t}$ can
have any value in the given range of $-1.5< \omega_{t}<1$.
\\
\\
{\bf Type 2}:
\\
Now we consider anisotropic fluid given in (\ref{e43}). Thus, the
coefficients in (\ref{AtoI}) become
\begin{eqnarray}\label{Ato4}
&&A=\frac{1}{12}k_{5}^{2},\nonumber\\
&&B=k_{4},\nonumber\\
&&C=\frac{1}{12}k_{5}^{2}(2p_{0r}p_{0t}-p_{0r}^2-p_{0t}^2),\nonumber\\
&&D=G = \frac{1}{12}k_{5}^{2}(1+2\omega),\nonumber\\
&&E=\frac{1}{6}k_{5}^{2}(\omega p_{0t}-\omega p_{0r}+p_{0t})+k_{4}\omega,\nonumber\\
&&F=\frac{1}{12}k_{5}^{2}(p_{0r}^2-p_{0t}^2)-k_{4}p_{0r},\nonumber\\
&&H=k_{4}\omega+\frac{1}{12}k_{5}^{2}(\omega p_{0r}-\omega p_{0t}+p_{0r}+p_{0t}),\nonumber\\
&&I=\frac{1}{12}k_{5}^{2}(p_{0r}p_{0t}-p_{0r}^2)-k_{4}p_{0t}.
\end{eqnarray}
Then the constraints (\ref{f10}) and (\ref{rho}) obtained for the
existence of a static state, respectively, become
\begin{eqnarray}\label{f44}
\Lambda -
\frac{1}{12}k_{5}^{2}(1+2\omega)\rho_{0}^{2}-\left(k_{4}\omega+\frac{1}{12}k_{5}^{2}(\omega
p_{0r}-\omega
p_{0t}+p_{0r}+p_{0t})\right)\rho_{0}+\frac{1}{12}k_{5}^{2}(p_{0r}p_{0t}-p_{0r}^2)-k_{4}p_{0t}=0,
\end{eqnarray}
and
\begin{eqnarray}\label{f45}
\frac{1}{6}k_{5}^{2}(1+\omega)\rho_{0}^{2}+\left(k_{4}(1+\omega)+\frac{1}{6}k_{5}^{2}(\omega
p_{0t}-\omega
p_{0r}+p_{0t})\right)\rho_{0}+\frac{1}{6}k_{5}^{2}(p_{0r}p_{0t}-p_{0r}^2)+k_{4}p_{0r}=0
.
\end{eqnarray}
In this case, the conditions (\ref{f25}) and (\ref{beta2}) governing
the stability of the the static state
become\begin{equation}\label{cond2}
\frac{\frac{1}{6}k_{5}^{2}(1+2\omega)\rho_{0}+\frac{1}{6}k_{5}^{2}(\omega
p_{0t}-\omega
p_{0r}+p_{0t})+k_{4}\omega}{\frac{1}{6}k_{5}^{2}\rho_{0}+k_4}<-1,
\end{equation}
and
\begin{equation}\label{beta4}
\rho_{0}\neq
-\frac{3k_{4}}{k_{5}^{2}}-\frac{1}{4}\left(p_{0r}+p_{0t}(\frac{1-\omega}{1+\omega})\right).
\end{equation}
Figure \ref{general2} shows the
possibility of satisfying the conditions (\ref{f44}) to
(\ref{beta4}). Similar to GR, considering $\omega< -1$, we can have
a stable static state for the given value of parameters.
 %%%%%%%%%%%%%%%%%%%%%%%%%%%%%%%%%%%%%%%%%%%%%%%%%%%%%%%%5
\subsubsection{Energy-momentum source with generalized Chaplygin gas
equation of state} 
\begin{figure}[ht]
\centering
\includegraphics[scale=0.55]{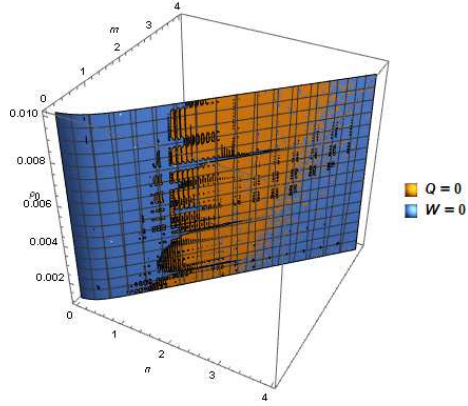}
\caption{\label{chaplygin} This figure
shows the surfaces which satisfy equations (\ref{c22}) and
(\ref{Chap}) in the solution space of equation (\ref{ba}). $Q=0$ and
$W=0$ represent equations (\ref{c22}) and (\ref{Chap}),
respectively, and their overlap shows the region in which we have a
stable static state. The used typical values are: $k_{4}=1$,
$k_{5}=1.5$, $\alpha_{r}=\frac{1}{2}$, $\alpha_{t}=2$,
$\Lambda=10^{-5}$.}
\end{figure}

For an energy-momentum source with a generalized
Chaplygin equation of state of the form (\ref{eos2}),  the Einstein
field equations on the brane will be
\begin{eqnarray}
&&\frac{\dot a_{2}^{2}}{a_{2}^{2}}+\frac{2\dot a_{1}\dot
a_{2}}{a_{1}a_{2}}+\frac{1}{a_{2}^{2}}=\Lambda+ k_{4}\rho+\frac{1}{12}k_{5}^{2}\left(\rho^2
-\frac{\alpha_{r}^{2}}{\rho^{2n}}-\frac{\alpha_{t}^{2}}{\rho^{2m}}+2\frac{\alpha_{r}\alpha_t}{\rho^{m+n}}\right),\label{c20}\\
&&\frac{2\ddot a_{2}}{a_{2}}+\frac{\dot
a_{2}^{2}}{a_{2}^{2}}+\frac{1}{a_{2}^{2}}=\Lambda+
k_{4}\frac{\alpha_{r}}{\rho^{n}} - \frac{1}{12}k_{5}^{2}\left(\rho^2
-\frac{\alpha_{r}^{2}}{\rho^{2n}}+\frac{\alpha_{t}^{2}}{\rho^{2m}}-2\frac{\alpha_t}{\rho^{m-1}}\right),\label{c21}\\
&&\frac{\ddot a_{1}}{a_{1}}+\frac{\ddot a_{2}}{a_{2}}+\frac{\dot
a_{1}\dot a_{2}}{a_{1}a_{2}}=\Lambda+
k_{4}\frac{\alpha_t}{\rho^m}-\frac{1}{12}k_{5}^{2}\left(\rho^2
+\frac{\alpha_{r}^{2}}{\rho^{2n}}-\frac{\alpha_{r}}{\rho^{n-1}}-\frac{\alpha_{t}}{\rho^{m-1}}-\frac{\alpha_{r}\alpha_t}{\rho^{m+n}}\right)\label{c22}.
\end{eqnarray}
The corresponding static state is given by the following equations
\begin{eqnarray}
&&\frac{1}{a_{02}^{2}}=\Lambda+
k_{4}\rho_0+\frac{1}{12}k_{5}^{2}\left(\rho_0^2
-\frac{\alpha_{r}^{2}}{\rho_0^{2n}}-\frac{\alpha_{t}^{2}}{\rho_0^{2m}}+2\frac{\alpha_{r}\alpha_t}{\rho_0^{m+n}}\right),\label{c200}\\
&&\frac{1}{a_{02}^{2}}=\Lambda+ k_{4}\frac{\alpha_{r}}{\rho_0^{n}} -
\frac{1}{12}k_{5}^{2}\left(\rho_0^2
-\frac{\alpha_{r}^{2}}{\rho_0^{2n}}+\frac{\alpha_{t}^{2}}{\rho_0^{2m}}-2\frac{\alpha_t}{\rho_0^{m-1}}\right),\label{c210}\\
&&0=\Lambda+
k_{4}\frac{\alpha_t}{\rho_0^m}-\frac{1}{12}k_{5}^{2}\left(\rho_0^2
+\frac{\alpha_{r}^{2}}{\rho_0^{2n}}-\frac{\alpha_{r}}{\rho_0^{n-1}}-\frac{\alpha_{t}}{\rho_0^{m-1}}-\frac{\alpha_{r}\alpha_t}{\rho_0^{m+n}}\right)\label{c220}.
\end{eqnarray}
Combining equations (\ref{c200}) and (\ref{c210}) we get
\begin{equation}\label{Chap}
k_{4}\left(\rho_{0}-\frac{\alpha_{r}}{\rho_0^n}\right)+\frac{1}{6}k_{5}^{2}\left(\rho_0^2
-\frac{\alpha_{r}^{2}}{\rho_0^{2n}}+\frac{\alpha_{r}\alpha_{t}}{\rho_0^{m+n}}-\frac{\alpha_{t}}{\rho_0^{m-1}}\right)=0
\end{equation}
Thus to have a static state the constrains (\ref{c220}) and
(\ref{Chap}) should be satisfied. Then, the perturbed field
equations versus the scalar perturbations (\ref{e16}) take the forms
\begin{eqnarray}\label{c23}
\frac{1}{a^{2}_{02}}-\frac{2\delta a_2}{a^{2}_{02}}&=&\Lambda+
k_{4}\rho_0(1+\delta\rho)\nonumber\\
&+&\frac{k_{5}^{2}}{12}\left(\rho_0^2(1+2\delta\rho)
-\frac{\alpha_{r}^{2}}{\rho_0^{2n}}(1-2n\delta\rho)-\frac{\alpha_{t}^{2}}
{\rho_0^{2m}}(1-2m\delta\rho)+2\frac{\alpha_{r}\alpha_t}{\rho_0^{m+n}}(1-(n+m)\delta\rho)\right),\\
\label{c24}
2\delta \ddot a_{2}+\frac{1}{a^{2}_{02}}-\frac{2\delta
a_2}{a^{2}_{02}}&=&\Lambda+k_{4}\frac{\alpha_{r}}{\rho_0^{n}}(1-n\delta\rho)\nonumber\\
&-& \frac{k_{5}^{2}}{12}\left(\rho_0^2(1+2\delta\rho)
-\frac{\alpha_{r}^{2}}{\rho_0^{2n}}(1-2n\delta\rho)
+\frac{\alpha_{t}^{2}}{\rho_0^{2m}}(1-2m\delta\rho)-\frac{2\alpha_t}{\rho_0^{m-1}}(1-(m-1)\delta\rho)\right),\\
\label{c25} \delta \ddot a_{1}+\delta \ddot a_{2}&=& \Lambda+
k_{4}\frac{\alpha_t}{\rho_0^m}(1-m\delta\rho)\nonumber\\
&-&\frac{k_{5}^{2}}{12}\left(\rho_0^2(1+2\delta\rho)
+\frac{\alpha_{r}^{2}}{\rho_0^{2n}}(1-2n\delta\rho)-\frac{\alpha_{r}}{\rho_0^{n-1}}(1-(n-1)\delta\rho\right)\nonumber\\
&-&\frac{\alpha_{t}}{\rho_0^{m-1}}(1-(m-1)\delta\rho)
-\frac{\alpha_{r}\alpha_t}{\rho_0^{m+n}}\left(1-(n+m)\delta\rho\right)).
\end{eqnarray}
Using equations (\ref{c200}) to (\ref{c220}) and defining
\begin{eqnarray}
&& \label{p1} A=k_{4}\rho_0+\frac{k_{5}^{2}}{6}\left(\rho_0^2
+n\frac{\alpha_{r}^{2}}{\rho_0^{2n}}+m\frac{\alpha_{t}^{2}}{\rho_0^{2m}}-(n+m)\frac{\alpha_{r}\alpha_t}{\rho_0^{m+n}}\right),\\
&&\label{p2} B=-nk_{4}\frac{\alpha_{r}}{\rho_0^{n}}-
\frac{k_{5}^{2}}{6}\left(\rho_0^2
+n\frac{\alpha_{r}^{2}}{\rho_0^{2n}}-m\frac{\alpha_{t}^{2}}{\rho_0^{2m}}+(m-1)\frac{\alpha_t}{\rho_0^{m-1}}\right),\\
&&\label{p3}
C=-mk_{4}\frac{\alpha_t}{\rho_0^m}-\frac{k_{5}^{2}}{12}\left(2\rho_0^2
-2n\frac{\alpha_{r}^{2}}{\rho_0^{2n}}+(n-1)\frac{\alpha_{r}}{\rho_0^{n-1}}
+(m-1)\frac{\alpha_{t}}{\rho_0^{m-1}}+(n+m)\frac{\alpha_{r}\alpha_t}{\rho_0^{m+n}}\right).
\end{eqnarray}
The perturbed equations (\ref{c23}) to (\ref{c25}) reduce to
\begin{eqnarray}
&&\label{c26} -\frac{2\delta a_2}{a^{2}_{02}}=
A\delta\rho,\\
&&\label{c27} 2\delta \ddot a_{2}-\frac{2\delta a_2}{a^{2}_{02}}=
B\delta\rho,\\
&&\label{c28} \delta \ddot a_{1}+\delta \ddot a_{2}=C\delta\rho.
\end{eqnarray}
Combining (\ref{c26}) and (\ref{c27}) we obtain\begin{equation}\label{c29}
\delta \ddot a_{2}+\left(\frac{B}{A}-1\right)\frac{1}{a_{02}^2}\delta a_2 =0.
\end{equation}
Then an  oscillatory mode for $\delta a_2$ is subjected to the
condition\begin{equation}\label{ba}\frac{B}{A}>1 .
\end{equation}
Combining equations (\ref{c26})-(\ref{c28}), we also obtain
\begin{equation}\label{c30}
-2\delta \ddot a_{1}=(-2C+B-A)\delta\rho.
\end{equation}
Using (\ref{c26}) and (\ref{c27})
\begin{equation}\label{c31}
2\delta \ddot a_{2}=(B-A)\delta\rho,
\end{equation}
Combining it with (\ref{c30}) we get
\begin{equation}\label{c32}
\delta \ddot a_{1}=\left(\frac{2C}{B-A}-1\right)\delta \ddot a_{2}.
\end{equation}
Then, the dynamics of $\delta a_1$ reads as
\begin{equation}\label{c320}
\delta  a_{1}=\left(\frac{2C}{B-A}-1\right)\delta a_{2}+\alpha
t+\beta.
\end{equation}
where $\alpha, \beta$ are constants and for $\alpha=\beta=0$ and
$B\neq A$ the oscillatory modes are possible and the static state
will be stable. Figure \ref{chaplygin} shows the possibility of
satisfying the constraints (\ref{c22}) and (\ref{Chap}) in the
solution space of equation (\ref{ba}) for the given values of
parameters. It is seen from the figure that we need $n,m>1$ and very
small amount of $\rho_{0}$, which corresponds to a large radial and
lateral pressure, to have a stable static state. However, for
$\rho_{0}<0.004$, $m$ can be less than 1.

Hence, the following is the summary of the analysis of the existence
and stability of a KSSS on a brane:\textit{ $(i)$ A perfect fluid
cannot support a finite size static KS geometry even in
the presence of higher dimensional modifications, $(ii)$ In contrast
to GR in four dimensions, an anisotropic fluid with
$p_{r}=\omega_{r}\rho$ and $p_{t}=\omega_{t}\rho~$ supports a stable
finite size static KS geometry, and $(iii)$ A
modification of the perfect fluid form as $p_{r}=\omega\rho +p_{0r}$
and $p_{t}=\omega\rho +p_{0t}$ can also support a stable nonsingular
KS type static state, and $(iv)$ In contrast to the
case in four dimensions, a stable nonsingular KS
geometry can be supported by a generalized Chaplygin gas source of
energy-momentum. }

\section{KS geometry in five dimensions}
In this section, we consider a five-dimensional KS type metric
\begin{equation}
ds^2=-dt^{2}+a_{1}^{2}(t)dr^{2}+a_{2}^{2}(t)(d\theta_{1}^{2}+sin^{2}\theta_{1}d\theta_{2}^{2}+sin^{2}\theta_{1}sin^{2}\theta_{2}d\theta_{3}^{2}),
\end{equation}
where $a_{1}(t)$ and $a_{2}(t)$ are two arbitrary functions of time
and the energy-momentum tensor supporting this geometry has the
generic form ${T^{\mu}}_{\nu}=diag\left(-\rho,\,p_r,\,p_t,
\,p_t,\,p_t\right)$. Then, the Einstein field equations are
\begin{eqnarray}
&&3(\frac{\dot a_{2}^{2}}{a_{2}^{2}}+\frac{\dot a_{1}\dot
a_{2}}{a_{1}a_{2}}+\frac{1}{a_{2}^{2}})=\Lambda+ k_{5}\rho,\label{fe11}\\
&&3(\frac{\ddot a_{2}}{a_{2}}+\frac{\dot
a_{2}^{2}}{a_{2}^{2}}+\frac{1}{a_{2}^{2}})=\Lambda- k_{5} p_{r},\label{fe22}\\
&&\frac{\ddot a_{1}}{a_{1}}+\frac{2\ddot a_{2}}{a_{2}}+\frac{2\dot
a_{1}\dot a_{2}}{a_{1}a_{2}}+\frac{\dot
a_{2}^{2}}{a_{2}^{2}}+\frac{1}{a_{2}^{2}}=\Lambda-
k_{5}p_{t}\label{fe33}.
\end{eqnarray}
\subsection{KSSS and stability analysis
in five dimensions} In the following, we study the existence and
stability of five-dimensional KS type static state considering the
field equations (\ref{fe11})-(\ref{fe33}). Similar to previous
sections, we consider two generic kinds of energy-momentum sources:
$(i)$ a fluid possessing linear equation of state, and $(ii)$ a
fluid with generalized Chaplygin gas type equations of state.
\subsubsection{Energy-momentum source with
linear equations of state} By considering a general equation of
state with the form given in (\ref{eos1}), the field equations
(\ref{fe11})-(\ref{fe33}) become
\begin{eqnarray}
&&3(\frac{\dot a_{2}^{2}}{a_{2}^{2}}+\frac{\dot a_{1}\dot
a_{2}}{a_{1}a_{2}}+\frac{1}{a_{2}^{2}})=\Lambda+ k_{5}\rho,\label{fe44}\\
&&3(\frac{\ddot a_{2}}{a_{2}}+\frac{\dot
a_{2}^{2}}{a_{2}^{2}}+\frac{1}{a_{2}^{2}})=\Lambda- k_{5}
\omega_{r}\rho -k_{5}p_{0r},\label{fe55}\\
&&\frac{\ddot a_{1}}{a_{1}}+\frac{2\ddot a_{2}}{a_{2}}+\frac{2\dot
a_{1}\dot a_{2}}{a_{1}a_{2}}+\frac{\dot
a_{2}^{2}}{a_{2}^{2}}+\frac{1}{a_{2}^{2}}=\Lambda-k_{5}\omega_{t}\rho
-k_{5} p_{0t}\label{fe66}.
\end{eqnarray}
Then, the corresponding static state is given by the following
equations
\begin{eqnarray}
&&\frac{3}{a_{02}^{2}}=\Lambda+ k_{5}\rho_{0},\label{e88}\\
&&\label{e99}
\frac{3}{a_{02}^{2}}=\Lambda-k_{5} \omega_{r}\rho_{0} -k_{5}p_{0r},\\
&&\label{e101} \frac{1}{a_{02}^{2}}=\Lambda-k_{5} \omega_{t}\rho_{0}
-k_{5} p_{0t}.
\end{eqnarray}
From (\ref{e88}) and (\ref{e99}) we get
\begin{equation}\label{e121}
p_{0r}=-(1+\omega_{r})\rho_{0}=\omega_{eff}\rho_{0},
\end{equation}
and combining (\ref{e88}) and (\ref{e101}) leads to
\begin{equation}\label{n2}
p_{0t}=\frac{2}{3}\frac{\Lambda}{k_{5}}-(\omega_t+\frac{1}{3})\rho_0.
\end{equation}
To study the stability of the static state given by
(\ref{e88})-(\ref{e101}), we consider the scalar perturbations in
the form of (\ref{e16}) and keep up to the first order perturbation
terms. Then equations (\ref{fe44})-(\ref{fe66}) give
\begin{eqnarray}\label{e171}
&&3(\frac{1}{a^{2}_{02}}-\frac{2\delta
a_2}{a^{2}_{02}})=\Lambda+k_{5}\rho_{0}+ k_{5}\rho_{0}\delta
\rho,\\
&&\label{e181} 3(\delta \ddot
a_{2}+\frac{1}{a^{2}_{02}}-\frac{2\delta a_2}{a^{2}_{02}})= \Lambda
-k_{5}\omega_{r}\rho_{0}\delta \rho-
k_{5}\omega_{r}\rho_{0}-k_{5}p_{0r},\\
&&\label{e191} \delta \ddot a_{1}+2\delta \ddot
a_{2}+\frac{1}{a^{2}_{02}}-\frac{2\delta a_2}{a^{2}_{02}}= \Lambda
-k_{5}\omega_{t}\rho_{0}\delta \rho-
k_{5}\omega_{t}\rho_{0}-k_{5}p_{0t}.
\end{eqnarray}
Using the static state defined in (\ref{e88})-(\ref{e101}), the
above equations reduce to
\begin{eqnarray}\label{e202}
&&-6\frac{\delta a_2}{a^{2}_{02}}= k_{5}\rho_{0}\delta
\rho,\\
&&\label{e212} 3\delta \ddot a_{2}-6\frac{\delta
a_2}{a^{2}_{02}}=-k_{5}\omega_{r}\rho_{0}\delta \rho,\\
&&\label{e222} \delta \ddot a_{1}+2\delta \ddot a_{2}-\frac{2\delta
a_2}{a^{2}_{02}}= -k_{5}\omega_{t}\rho_{0}\delta \rho.
\end{eqnarray}
Substituting  (\ref{e202})  in  (\ref{e212}) leads to
\begin{equation}\label{e242}
\delta \ddot a_{2}+\gamma^{2}\delta a_2 =0,
\end{equation}
where
\begin{equation}\label{ee282}
\gamma^{2}=-\frac{2(1+\omega_{r})}{a_{02}^{2}}=\frac{2
\omega_{eff}}{a_{02}^{2}}.
\end{equation}
Hence, the oscillating modes for $\delta a_2$
\begin{eqnarray}\label{e283}
\delta a_{2}=C_{1}e^{i\gamma t}+C_{2}e^{-i\gamma t},
\end{eqnarray}
requires the constraint\begin{equation}\label{e252} \omega_{r}<-1.
\end{equation}
Similarly, using (\ref{e202}), (\ref{e212}) and (\ref{e222}), we
obtain
\begin{equation}\label{e292}
\delta \ddot
a_{1}=k_{5}(\frac{2}{3}\omega_{r}-\omega_{t}+\frac{1}{3})\rho_{0}\delta\rho.
\end{equation}
Combining  (\ref{e202}) and (\ref{e212}) gives
\begin{equation}\label{e313}
\delta\rho=\frac{-3\delta \ddot a_{2}}{k_{5}(1+\omega_{r})\rho_{0}},
\end{equation}
where substituting in (\ref{e292})
yields\begin{equation}\label{e323} \delta \ddot
a_{1}=\left(\frac{3\omega_{t}-\omega_{r}}{1+\omega_{r}}-1\right)\delta
\ddot a_{2},
\end{equation}
Twice integration of this equation gives
\begin{eqnarray}\label{e333}
\delta
a_{1}=\left(\frac{3\omega_{t}-\omega_{r}}{1+\omega_{r}}-1\right)\delta
a_{2}+ C_{3}t+C_{4}.
\end{eqnarray}
Similar to 4D case, the stable oscillatory modes in $\delta a_1$ is
subjected to the condition $C3 = C4 = 0$.  Here, for
$3\omega_{t}=3\omega_{r}+2$ the perturbation amplitude on the radial
and lateral directions are the same.
\\
\\
In the following,  we consider two specific forms of the fluid
(\ref{eos1}) and discuss on the stability of KSSS.
\\
\\
{\bf 4.1.1.1 Perfect fluid}

To study perfect fluid, we set $\omega_{r}=\omega_{t}=\omega$ and
$p_{0r}=p_{0t}$. Then, equations (\ref{e88})-(\ref{e101}) for the
static state reduce to
 \begin{eqnarray}
&&\label{e343}
\frac{3}{a_{02}^{2}}=\Lambda+ k_{5}\rho_{0},\\
&&\label{e353}
\frac{3}{a_{02}^{2}}=\Lambda-k_{5}\omega\rho_{0},\\
&&\label{e363} \frac{1}{a_{02}^{2}}=\Lambda-k_{5} \omega\rho_{0}.
\end{eqnarray}
Comparing (\ref{e343}) and (\ref{e353}) leads to
$\omega=-1$. Similar to the case in four dimensions, (\ref{e363}) is not consistent with (\ref{e353}) regardless of $\omega$ values meaning that having
a static state is not possible for a perfect fluid in five dimensions.
\\
\\
{\bf 4.1.1.2 Anisotropic fluid}
\\
Here, similar to previous sections, we consider two modifications of
the perfect fluid.
\\
\\
{\bf Type 1}:

In this case, by considering two different equations of state
parameters for the radial and lateral directions in the form of
(\ref{e38}), the equations (\ref{e88})-(\ref{e101}) governing the
static state become
\begin{eqnarray} &&\label{e393}
\frac{3}{a_{02}^{2}}=\Lambda+ k_{5}\rho_{0},\\
&&\label{e404}
\frac{3}{a_{02}^{2}}=\Lambda-k_{5}\omega_{r}\rho_{0},\\
&&\label{e414} \frac{1}{a_{02}^{2}}=\Lambda-k_{5}\omega_{t}\rho_{0}.
\end{eqnarray}
comparing equations (\ref{e393}) and (\ref{e404}) we obtains the
constraint $\omega_{r}=-1$ which is not allowed based on equation
(\ref{e252}).
Then a Type 1 fluid fails to support a stable static sate.\\
\\
{\bf Type 2}:
\\
Now we consider anisotropic fluid with the form of (\ref{e43}). Then
the equations (\ref{e121}) and (\ref{n2}) for $p_{0r}$ and $p_{0t}$
become
\begin{equation}\label{n6}
p_{0r}=-(1+\omega)\rho_{0},
\end{equation}
\begin{equation}\label{n7}
p_{0t}=\frac{2}{3}\frac{\Lambda}{k_{5}}-(\omega+\frac{1}{3})\rho_0.
\end{equation}
In this case, $\gamma^2$ given by equation (\ref{ee282}) becomes
\begin{eqnarray}\label{eee28}
\gamma^{2}=-\frac{3(1+\omega)}{a_{02}^{2}},
\end{eqnarray}
Therefore, the positivity condition on $\gamma^2$ to have a stable
nonsingular static state demands
 \begin{equation}\label{omega}
 \omega<-1,
 \end{equation}
 which means that the fluid supporting the geometry lies in the phantom
 range. Also, equation (\ref{e323}) becomes
 \begin{equation}\label{ee323}
\delta \ddot a_{1}=\left(\frac{\omega-1}{\omega+1}\right)\delta
\ddot a_{2},
\end{equation}
The dynamics of $\delta a_1$ reads as
\begin{equation}\label{e474}
\delta a_{1}=\left(\frac{\omega -1}{\omega+1}\right)\delta a_{2}+
C_{3}t+C_{4}.
\end{equation}
Then, similar to four dimension, for $C3 = C4 = 0$ the
oscillatory modes are possible and hence the static state will be
stable.
\subsection{Energy-momentum source with generalized Chaplygin gas
equation of state} Using energy-momentum source with a generalized
Chaplygin equation of state of the form (\ref{eos2}),  the Einstein
field equations in five dimensions will be
\begin{eqnarray}
&&3(\frac{\dot a_{2}^{2}}{a_{2}^{2}}+\frac{\dot a_{1}\dot
a_{2}}{a_{1}a_{2}}+\frac{1}{a_{2}^{2}})=\Lambda+ k_{5}\rho,\label{n8}\\
&&3(\frac{\ddot a_{2}}{a_{2}}+\frac{\dot
a_{2}^{2}}{a_{2}^{2}}+\frac{1}{a_{2}^{2}})=\Lambda+k_{5}\frac{\alpha_r}{\rho^n},\label{n9}\\
&&\frac{\ddot a_{1}}{a_{1}}+\frac{2\ddot a_{2}}{a_{2}}+\frac{2\dot
a_{1}\dot a_{2}}{a_{1}a_{2}}+\frac{\dot
a_{2}^{2}}{a_{2}^{2}}+\frac{1}{a_{2}^{2}}=\Lambda+
k_{5}\frac{\alpha_t}{\rho^m}\label{n10}.
\end{eqnarray}
Then, the corresponding static state is
\begin{eqnarray}
&&\frac{3}{a_{02}^{2}}=\Lambda+ k_{5}\rho_{0},\label{n11}\\
&&\label{n12}
\frac{3}{a_{02}^{2}}=\Lambda+k_{5}\frac{\alpha_r}{\rho_0^n},\\
&&\label{n13}
\frac{1}{a_{02}^{2}}=\Lambda+k_{5}\frac{\alpha_t}{\rho_0^m}.
\end{eqnarray}
applying the perturbations given by (\ref{e16}) to the field
equations (\ref{n8})-(\ref{n10}) leads to
\begin{eqnarray}\label{n14}
&&3(\frac{1}{a^{2}_{02}}-\frac{2\delta
a_2}{a^{2}_{02}})=\Lambda+k_{5}\rho_{0}+ k_{5}\rho_{0}\delta
\rho,\\
&&\label{n15} 3(\delta \ddot
a_{2}+\frac{1}{a^{2}_{02}}-\frac{2\delta a_2}{a^{2}_{02}})= \Lambda
+\frac{k_5\alpha_r}{\rho_0^n}(1-n\delta\rho),\\
&&\label{n16} \delta \ddot a_{1}+2\delta \ddot
a_{2}+\frac{1}{a^{2}_{02}}-\frac{2\delta a_2}{a^{2}_{02}}=
\Lambda+\frac{k_5\alpha_t}{\rho_0^m}(1-m\delta\rho).
\end{eqnarray}
Considering equations (\ref{n11})-(\ref{n13}), the above equations
reduce to
\begin{eqnarray}\label{c131}
&&\frac{6\delta a_2}{a^{2}_{02}}= -k_{5}\rho_{0}\delta
\rho,\\
&&\label{c141} 3\delta \ddot a_{2}-\frac{6\delta
a_2}{a^{2}_{02}}= -nk_5\frac{\alpha_r}{\rho_0^n}\delta\rho,\\
&&\label{c151} \delta \ddot a_{1}+2\delta \ddot a_{2}-\frac{2\delta
a_2}{a^{2}_{02}}=-mk_{5}\frac{\alpha_t}{\rho_0^m}\delta\rho.
\end{eqnarray}
Combining (\ref{c131}) and (\ref{c141}) one gets
\begin{equation}\label{c161}
\delta \ddot a_{2}+\gamma^{2}\delta a_2 =0,
\end{equation}
where
\begin{equation}\label{gam11}
\gamma^{2}=-\frac{2(n\alpha_r \rho_0^{-(n+1)}+1)}{a_{02}^{2}}.
\end{equation}
Thus, $\gamma^2$ is always negative and consequently there are no
oscillatory modes for $\delta a_1$ and $\delta a_2$.

The summary of the result obtained in this section is as follows. \textit{The analysis here reveals that the existence and stability conditions for
a four and a five dimensional KS geometries without a brane are similar in some manners. More
specifically, (i) finite size static KS geometry  does not exist for a perfect
fluid source, (ii) an anisotropic type 1 fluid cannot support a static state,
but an anisotropic type 2 fluid supports a stable nonsingular KS type static
state, and (iii) a generalized Chaplygin gas fluid cannot  support a stable
nonsingular KS geometry in a 5-dimensional  model without brane. The results  of the analysis
are different than the case when a four dimensional
brane is embedded in a five (or higher) dimensional Ricci flat bulk space. Specifically, a stable nonsingular KS geometry can be supported by both the generalized Chaplygin gas fluid and an anisotropic fluid in a brane model. One
interpretation of the differences in these two 5-dimensional models
(with and without brane) is that in a braneworld scenario, the matter fields
are confined to the brane and have no way to propagate along the extra
dimension(s). Due to this confinement, matter fields have one less degree
of freedom in comparison to the case where they are distributed in a five
dimensional space. The confinement of the matter fields to the brane affects
the local extrinsic curvature and dynamics of the brane within its bulk space. This induces a modification
to the Einstein's field equations on the brane. This modification  provides
a geometrical interpretation for dark energy  as the manifestation of the local extrinsic shape of the brane, see
for instances \cite{1}.
In our study, this modification provides the possibility of the existence
and stability of an anisotropic KS type static state for a wider range of fluid
types on the brane in comparison to the four and five dimensional models without
brane. }

\section{ Conclusion}
In the present work,   the possibility of having a nonsingular  KS type spacetime
as a seed for an emergent universe is investigated.  It is discussed that the
existence and stability of the nonsingular KSSS depend on the  dimensions
of the spacetime and the nature of
the fluid supporting the geometry. In particular, it is found that:
\begin{itemize}
\item In the context of  GR in four dimensions:
\begin{description}
\item[$(i)$] A perfect
fluid cannot support a finite size static KS geometry.
\item[$(ii)$]
An anisotropic fluid with equations of states  $p_{r}=\omega_{r}\rho$ and $p_{t}=\omega_{t}\rho~$
can support a finite size static KS geometry but it is not stable
against the scalar perturbations.
\item[$(iii)$] A modification of the perfect
fluid form possessing equations of state $p_{r}=\omega\rho +p_{0r}$ and $p_{t}=\omega\rho +p_{0t}$ can
support a stable nonsingular KS type static state.
\item[$(iv)$] A generalized Chaplygin gas fluid with the equations of state
$p_{r}=-\frac{\alpha_r}{\rho^n}$ and $p_{t}=-\frac{\alpha_t}{\rho^m}$ cannot
support a stable nonsingular KS geometry.

\end{description}
\item In the context of a  five dimensional braneworld scenario:
\begin{description}
\item[$(i)$] A perfect
fluid cannot support a finite size static KS geometry even in
the presence of higher dimensional modifications.
\item[$(ii)$]
In contrast to  the four dimensional case, an anisotropic fluid having equations
of state $p_{r}=\omega_{r}\rho$ and $p_{t}=\omega_{t}\rho~$
supports a stable finite size static KS geometry.
\item[$(iii)$] A fluid having the equations of state $p_{r}=\omega\rho +p_{0r}$ and $p_{t}=\omega\rho +p_{0t}$ can
support a stable nonsingular KS type static state.
\item[$(iv)$] In contrast to the case in four dimensions, a stable nonsingular
KS geometry can be supported by a generalized Chaplygin gas fluid.
\end{description}
\item In the context of a  five dimensional model without brane, the results of the analysis for the existence and stability
conditions are similar to the four dimensional model addressed above.
\begin{description}
\item
\end{description}
\end{itemize}
%%%%%%%%%%%%%%%%%%%%%%%%%%%%%%%%%%%%%%%%%%%%%%%%%%%%%%%%%%%%%%%%%%%%%%%%%

\end{document}